\documentclass{elsart}

\usepackage{cite}
\usepackage{graphicx,amssymb}
\input{Qcircuit.tex}
\makeatletter
\@addtoreset{equation}{section}
\renewcommand{\theequation}{\thesection.\@arabic\c@equation}
\makeatother
\numberwithin{equation}{subsection}
\include{Qcircuit.tex}
\begin{document}

\begin{frontmatter}

\title{Synthesis of multi-qudit Hybrid and $d$-valued Quantum Logic Circuits by Decomposition}

\author[PsuMath]{Faisal Shah Khan\corauthref{cor}},
\corauth[cor]{Corresponding author.}
\ead{faisal@pdx.edu}
\author[PsuEce]{Marek Perkowski}
\ead{mperkows@ece.pdx.edu}
\address[PsuMath]{Portland State University, Department of Mathematics and Statistics,
Portland, Oregon 97207-0751, USA}
\address[PsuEce]{Portland State University, Department of Electrical and Computer Engineering,
Portland, Oregon 97207-0751, USA}

\begin{abstract}
Recent research in generalizing quantum computation from 2-valued qudits to $d$-valued qudits has shown practical advantages for scaling up a quantum computer. A further generalization leads to quantum computing with \emph{hybrid} qudits where two or more qudits have different finite dimensions. Advantages of hybrid and $d$-valued gates (circuits) and their physical realizations have been studied in detail by Muthukrishnan and Stroud (Physical Review A, 052309, 2000), Daboul et al. (J. Phys. A: Math. Gen. 36 2525-2536, 2003), and Bartlett et al (Physical Review A, Vol.65, 052316, 2002). In both cases, a quantum computation is performed when a unitary evolution operator, acting as a quantum logic gate, transforms the state of qudits in a quantum system. Unitary operators can be represented by square unitary matrices. If the system consists of a single qudit, then Tilma et al (J. Phys. A: Math. Gen. 35 (2002) 10467-10501) have shown that the unitary evolution matrix (gate) can be synthesized in terms of its Euler angle parametrization. However, if the quantum system consists of multiple qudits, then a gate may be synthesized by matrix decomposition techniques such as QR factorization and the Cosine-sine Decomposition (CSD). In this article, we present a CSD based synthesis method for \emph{n} qudit hybrid quantum gates, and as a consequence, derive a CSD based synthesis method for \emph{n} qudit gates where all the qudits have the same dimension.
\end{abstract}

\end{frontmatter}

\section{Introduction}
A \emph{qudit} replaces a classical dit as an information unit in $d$-valued quantum computing. A qudit is represented as a unit vector in the state space, which is a complex projective $d$ dimensional Hilbert space, $\mathcal{H}_{d}$.
In the computational basis, the basis vectors of $\mathcal{H}_{d}$ are written in Dirac notation as $\ket{0},\ket{1},
\dots \ket{d-1}$, where $\ket{i}=(0,0,\dots,1, \dots, 0)^{T}$ with a 1 in the $(i+1)$-st coordinate, for $0 \leq i \leq (d-1)$. An arbitrary vector $\ket{a}$ in $\mathcal{H}_{d}$ can be expressed as a linear combination $\ket{a}=\sum_{i=0}^{d-1}{x_{i}\ket{i}}$, $x_{i} \in \textbf{C}$ and $\sum{\left|x_{i}\right|^{2}}=1$. The real number $\left|x_{i}\right|^{2}$ is the probability that the state vector $\ket{a}$ will be in  $i$-th basis state upon measurement. 

When the state spaces of $n$ qudits of different $d$-valued dimensions are combined via their algebraic tensor product, the result is a $n$ qudit \emph{hybrid} state space $\mathcal{H}=\mathcal{H}_{d_{1}} \otimes \mathcal{H}_{d_{2}} \otimes \dots \otimes \mathcal{H}_{d_{n}}$, where $\mathcal{H}_{d_{i}}$ is the state space of the $d_{i}$-valued qudit. The computational basis for $\mathcal{H}$ would consist of all possible tensor products of the computational basis vectors of the component state spaces $\mathcal{H}_{d_{i}}$. If $d_i=d$ for each $i$, the resulting state space $\mathcal{H}_{d}^{\otimes{n}}$ is that of $n$ $d$-valued qudits. 

The \emph{evolution} of state space changes the state of the qudits via the action of a unitary operator on the qudits. A unitary operator can be represented by a unitary evolution matrix. For the hybrid state space $\mathcal{H}$, an evolution matrix will have size $(d_{1}d_{2} \ldots d_{N}) \times (d_{1} d_{2} \ldots d_{N})$, while the evolution matrix for $\mathcal{H}_{d}^{\otimes{n}}$ will be of size $d^{n} \times d^{n}$. In the context of quantum logic synthesis, an evolution matrix is a quantum logic circuit that needs to be realized by a universal set of quantum logic gates. It is well established that sets of one and two qudit quantum gates are universal~\cite{Brylinski:02,Daboul:02, MuthuStroud:04,Tilma:02}. Hence, the synthesis of an evolution matrix requires that the matrix be decomposed to the level of unitary matrices acting on one or two qudits.

Unitary matrix decomposition methods like the QR factorization and the Cosine Sine decomposition from matrix perturbation theory have been used for 2-valued and 3-valued quantum logic synthesis. In these domains, qudits are referred to as \emph{qubits} and \emph{qutrits} respectively. The Cosine Sine decomposition (CSD) of a unitary matrix, discussed in section~\ref{sect:CSD}, has been used by M\"ott\"onen et. al~\cite{mottonen:04} and Shende et. al~\cite{shende:05} to iteratively synthesize multi-qubit quantum circuits. The authors of this article recently extended the CSD to iterated synthesis of 3-valued quantum logic circuits acting on $n$ qutrits~\cite{FSK:05}. Bullock et.al have recently presented a synthesis method for $n$ qudit quantum logic gates using a variation of the QR matrix factorization~\cite{bullock:04}. This article presents a CSD based method for synthesis of $n$ qudit hybrid and $d$-valued quantum logic gates.

\section{The Cosine-Sine Decomposition (CSD)}\label{sect:CSD}
Let the unitary matrix $\textit{W}\in \textbf{C}^{m\times m}$ be partitioned in $2 \times 2$ block form as
\begin{equation}\label{eqn:CSD matrix}
W=\bordermatrix {  &r      & m-r    \cr
                 r &W_{11} & W_{12} \cr
               m-r &W_{21} & W_{22} \cr}
\end{equation}
with $2r\leq m$. Then there exist $r \times r$ unitary
matrices $U$ and $X$, $r \times r$ real diagonal matrices $C$ and
$S$, and $(m-r) \times (m-r)$ unitary matrices $V$ and $Y$ such that
\begin{equation}\label{eqn:CSD1}
W  = \left(\begin{array}{cc}
U & 0 \\ 0 & V
\end{array}\right)
\left(\begin{array}{ccc}
  C & -S & 0 \\ S & C & 0 \\ 0 & 0 & I_{m-2r}
\end{array}\right)
\left(\begin{array}{cc}
 X & 0 \\ 0 & Y
\end{array}\right)
\end{equation}
The matrices $C$ and $S$ are the so-called cosine-sine matrices and are of the form $C$ = diag$(\cos \theta_{1}, \cos\theta_{2}, \dots,\cos\theta_{r})$, $S$ = diag$(\sin \theta_{1}$, $\sin \theta_{2},\dots,\sin \theta_{r})$ such that $\sin^{2}\theta_{i}+\cos^{2}\theta_{i}=1$ for some $\theta_{i}$, $1 \leq i \leq r$~\cite{Stewart:90}. Algorithms for computing the CSD and the angles $\theta_{i}$ are given in~\cite{Bjorck:73, Stewart:82}.
The CSD is essentially the well known singular value decomposition of a unitary matrix implemented at the block matrix level~\cite{Paige:92}. In sections~\ref{sect:2-valued CSD} and~\ref{sect:3-valued CSD}, we give an overview of the CSD based synthesis methods of 2 and 3-valued quantum logic circuits, respectively. From now on, we will not distinguish between gates, circuits and their corresponding unitary matrices. 

\section{Synthesis of 2-valued (binary) Quantum Logic Circuits}\label{sect:2-valued CSD}
As shown in~\cite{FSK:05, mottonen:04,shende:05,Tucci:98}, the CS decomposition gives a recursive method for synthesizing 2-valued and 3-valued $n$ qudit quantum logic gates. In the 2-valued case the CSD of a $2^{n} \times 2^{n}$ unitary matrix $W$ reduces to the form
\begin{equation}\label{eqn:CSD2}
W  = \left(\begin{array}{cc}
  U & 0 \\ 0 & V
\end{array}\right)
\left(\begin{array}{cc}
  C & -S \\ S & C \\
\end{array}\right)
\left(\begin{array}{cc}
  X & 0 \\  0 & Y
\end{array}\right)
\end{equation}
with each block matrix in the decomposition of size $2^{n-1} \times 2^{n-1}$. 

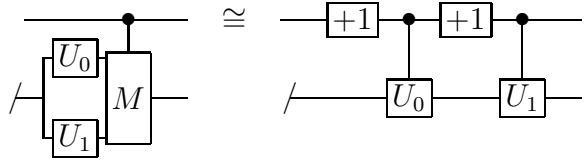
\begin{figure}\centerline{
 \Qcircuit @C.30em @R=.1em {
& &\qw &\qw  &\qw &\ctrl{1} &\qw &\qw &\qw &\qw & & & & &\cong & & & & & &\qw &\qw &\qw &\qw &\gate{+1} &\ctrl{2} &\gate{+1} &\ctrl{2} &\qw &\qw &\qw &\qw \\
& & & &\gate{U_{0}} &\multigate{2}{M} & & & & & & & & & & & \\
/ &\qw &\qw &\qw\qwx & &\gost{M} &\qw &\qw &\qw &\qw & & & & & & & & & & &/ &\qw &\qw &\qw &\qw &\gate{U_{0}} &\qw &\gate{U_{1}} &\qw &\qw &\qw &\qw\\
& & &\qwx &\gate{U_{1}} &\ghost{M} \\}}
\caption{\scriptsize{2-valued Quantum Multiplexer $M$ controlling the lower $(n-1)$ qubits by the top qubit. The slash symbol (/) represents $(n-1)$ qubits on the second wire. The gates labeled +1 are shifters (inverters in 2-valued logic), increasing the value of the qubit by 1 mod 2 thereby allowing for control by the highest qubit value. Depending on the value of the top qubit, one of $U_{t}$ is applied to the lower qubits for $t\in\left\{0,1\right\}$.}}\label{fig:2-valued QMUX}
\end{figure}

In terms of synthesis, the block diagonal matrices in (\ref{eqn:CSD2}) are \emph{quantum multiplexers}~\cite{shende:05}. A quantum multiplexer is a gate acting on $n$ qubits of which one is designated as the control qubit. If the control qubit is the highest order qubit, the multiplexer matrix is block diagonal. Depending on whether the control qubit carries $\ket{0}$ or $\ket{1}$, the gate then performs either the top left block or the bottom right block of the $n \times n$ block diagonal matrix on the remaining $(n-1)$ qubits. A circuit diagram for a $n$ qubit quantum multiplexer with the highest order control qubit is given in figure~\ref{fig:2-valued QMUX}. Observe that we decomposed arbitrary quantum multiplexer to single qubit gates and $n$ qubit standard controlled gates. The controlled gates execute the operator in the box when the controlling qubit has values 1 (mod 2). Such a quantum multiplexer can be expressed as
\begin{equation}\label{eqn:n qubit qmux}
\left(\begin{array}{cc} U_{0} & 0 \\ 0 & U_{1}
\end{array}\right)
\left(\ket{a_{1}}\otimes \ket{a_{2}} \otimes \dots \otimes \ket{a_{n}}\right)
\end{equation}
where $\ket{a_{i}}$ is the $i$-th qubit in the circuit, and both block matrices $U_{0}$ and $U_{1}$ are of size $2^{n-1} \times 2^{n-1}$. Depending on whether $\ket{a_{1}}=\ket{0}$ or $\ket{a_{1}}=\ket{1}$, the expression (\ref{eqn:n qubit qmux}) reduces to 
\begin{equation}\label{eqn:n qubit qmux1}
\ket{0} \otimes 
U_{0}
\left(\ket{a_{2}} \otimes \ket{a_{3}} \otimes \dots \otimes \ket{a_{n}} \right)
\end{equation}
or 
\begin{equation}\label{eqn:n qubit qmux2}
\ket{1} \otimes 
U_{1}
\left(\ket{a_{2}} \otimes \ket{a_{3}} \otimes \dots \otimes \ket{a_{n}}\right)
\end{equation}
respectively.

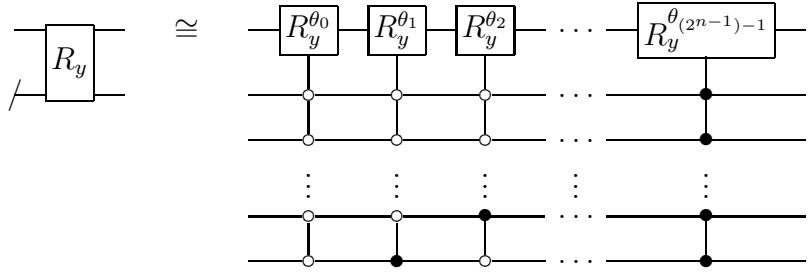
\begin{figure}\centerline{
 \Qcircuit @C=1em @R=1em {
& &\multigate{1}{R_{y}} &\qw & &\cong & & &\gate{R_{y}^{\theta_{0}}} &\gate{R_{y}^{\theta_{1}}} &\gate{R_{y}^{\theta_{2}}} &\qw &\dots & &\gate{R_{y}^{\theta_{(2^{n-1})-1}}} &\qw \\
& /&\ghost{R_{y}} &\qw & & & & &\ctrlo{-1} &\ctrlo{-1} &\ctrlo{-1} &\qw &\dots & &\ctrl{-1} &\qw  \\
& & & & & & & &\ctrlo{-1}  &\ctrlo{-1} &\ctrlo{-1} &\qw &\dots & &\ctrl{-1} &\qw \\
& & & & & & & &\vdots &\vdots &\vdots & &\vdots & &\vdots\\
& & & & & & & &\ctrlo{1} &\ctrlo{1} &\ctrl{1} &\qw &\dots & &\ctrl{1} &\qw\\
& & & & & & & &\ctrlo{-1} &\ctrl{-1} &\ctrlo{-1} &\qw &\dots & &\ctrl{-1} &\qw\\ }}
\caption{\scriptsize{A uniformly $(n-1)$-controlled $R_{y}$ rotation for 2-valued quantum logic. The lower $(n-1)$ qubits are the control qubits represented on the left hand side  by the symbol / on the second wire. The $\circ$ control turns on for control value $\ket{0}$ and the $\bullet$ control turns on for control value $\ket{1}$. It requires $2^{n-1}$ one qubit controlled gates $R_{y}^{\theta_{i}}$ to implement a uniformly $(n-1)$-controlled $R_{y}$ rotation. }}\label{fig:(n-1)-unifo control Ry}
\end{figure}

The cosine-sine matrix in (\ref{eqn:CSD2}) is realized as a \emph{uniformly $(n-1)$-controlled $R_{y}$ rotation gate}, a variation of the quantum multiplexer. As shown in figure~\ref{fig:(n-1)-unifo control Ry}, a uniformly $(n-1)$-controlled $R_{y}$ rotation gate $R_{y}$ is composed of a sequence of $({n-1})$-fold controlled gates $R_{y}^{\theta_{i}}$, all acting on the highest order qubit, where
\begin{equation}\label{eqn:R_{y}2}
R_{y}^{\theta_{i}}  = \left(\begin{array}{cc}
  \cos\theta_{i} & -\sin\theta_{i} \\  \sin\theta_{i} &  \cos\theta_{i}
\end{array}\right).
\end{equation}

The control selecting the angle $\theta_{i}$ in the gate $R_{y}^{\theta_{i}}$ depends on which of the $({n-1})$ basis state configurations the control qubits are in at that particular stage in the circuit. In figure~\ref{fig:(n-1)-unifo control Ry}, the open controls represent the basis state $\ket{0}$ and a filled in control represents basis state $\ket{1}$. The $i$-th $(n-1)$-controlled gate $R_{y}^{\theta_{i}}$ may be expressed as
\begin{equation}\label{eqn:n qubit uniformly controlled R_y}
\left(\begin{array}{cc} \cos\theta_{i} & -\sin\theta_{i} \\ \sin\theta_{i} &\cos\theta_{i}
\end{array}\right)
\ket{a_{1}} \otimes \left(\ket{a_{2}} \otimes \dots \otimes \ket{a_{n}}\right)
\end{equation}
with $\theta_{i}$ taking on values from the set $\left\{\theta_{0}, \theta_{1}, \dots, \theta_{2^{n-1}-1}\right\}$ depending on the configuration of $\left(\ket{a_{2}} \otimes \dots \otimes \ket{a_{n}}\right)$, resulting in a specific $R_{y}^{\theta_{i}}$ for each $i$.

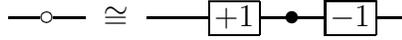
\begin{figure}\centerline{
 \Qcircuit @C1em @R=.8em {
&\controloo &\qw &\cong & &\qw &\gate{+1}  &\controll &\gate{-1} &\qw  }}
\caption{\scriptsize{A control by input value 0 (mod 2) realized in terms of control by the highest value 1 (mod 2).}}\label{fig:0-control}
\end{figure}

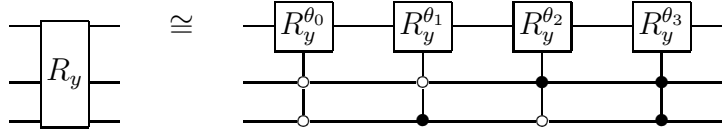
\begin{figure}\centerline{
 \Qcircuit @C1em @R=.8em {
&\multigate{2}{R_{y}} &\qw & &\cong & &  &\gate{R_{y}^{\theta_{0}}} &\qw &\gate{R_{y}^{\theta_{1}}} &\qw &\gate{R_{y}^{\theta_{2}}} &\qw &\gate{R_{y}^{\theta_{3}}} &\qw \\
&\ghost{R_{y}} &\qw & & & &  &\ctrlo{-1} &\qw &\ctrlo{-1} &\qw &\ctrl{-1} &\qw &\ctrl{-1} &\qw\\
&\ghost{R_{y}} &\qw & & & &  &\ctrlo{-1} &\qw &\ctrl{-1} &\qw &\ctrlo{-1} &\qw &\ctrl{-1}  &\qw\\}}
\caption{\scriptsize{A uniformly 2-controlled $R_{y}$ rotation in 2-valued logic: the lower two qubits are the control qubits,
 and the top bit is the target bit.}}\label{fig:2-unifo control Ry}
\end{figure}

As an example, consider the 3 qubit uniformly 2-controlled $R_{y}$ gate controlling the top qubit from figure~\ref{fig:2-unifo control Ry}. Then the action of $R_{y}^{\theta_{i}}$ on the circuit is 
\begin{equation}\label{eqn:n qubit uniformly controlled R_y1}
\left(\begin{array}{cc} \cos\theta_{i} & -\sin\theta_{i} \\ \sin\theta_{i} &\cos\theta_{i}
\end{array}\right)
\ket{a_{1}} \otimes
\left(\ket{a_{2}} \otimes \ket{a_{3}}\right)
\end{equation}
with $\theta_{i} \in \left\{\theta_{0}, \theta_{1}, \theta_{2}, \theta_{3}\right\}$. As $\ket{a_{2}} \otimes \ket{a_{3}}$ takes on the values from the set \\ $\left\{\ket{0}\otimes \ket{0}, \ket{0}\otimes \ket{1}, \ket{1}\otimes \ket{0}, \ket{1}\otimes \ket{1}\right\}$ in order, the expression in (\ref{eqn:n qubit uniformly controlled R_y1}) reduces to the following 4 expressions respectively.

\begin{equation}\label{eqn:n qubit uniformly controlled R_y2}
\left(\begin{array}{cc} \cos\theta_{0} & -\sin\theta_{0} \\ \sin\theta_{0} &\cos\theta_{0}
\end{array}\right)
\ket{a_{1}} \otimes
\left(\ket{0} \otimes \ket{0}\right)
\end{equation}
\begin{equation}\label{eqn:n qubit uniformly controlled R_y3}
\left(\begin{array}{cc} \cos\theta_{1} & -\sin\theta_{1} \\ \sin\theta_{1} &\cos\theta_{1}
\end{array}\right)
\ket{a_{1}} \otimes
\left(\ket{0} \otimes \ket{1}\right)
\end{equation}
\begin{equation}\label{eqn:n qubit uniformly controlled R_y4}
\left(\begin{array}{cc} \cos\theta_{2} & -\sin\theta_{2} \\ \sin\theta_{2} &\cos\theta_{2}
\end{array}\right)
\ket{a_{1}} \otimes
\left(\ket{1} \otimes \ket{0}\right)
\end{equation}
\begin{equation}\label{eqn:n qubit uniformly controlled R_y5}
\left(\begin{array}{cc} \cos\theta_{3} & -\sin\theta_{3} \\ \sin\theta_{3} &\cos\theta_{3}
\end{array}\right)
\ket{a_{1}} \otimes
\left(\ket{1} \otimes \ket{1}\right)
\end{equation}

\section{CSD Synthesis of 3-valued (ternary) Quantum Logic Circuits}\label{sect:3-valued CSD}
In the 3-valued case, two applications of the CSD are needed to decompose a $3^{n} \times 3^{n}$ unitary matrix $W$ to the point where every block in the decomposition has size $3^{n-1} \times 3^{n-1}$~\cite{FSK:05}. Choose the parameters $m$ and $r$ given in (\ref{eqn:CSD matrix}) as $m=3^{n}$ and $r=3^{n-1}$, so that $m-r = 3^{n}-3^{n-1}=3^{n-1}(3-1)=3^{n-1}\cdot2$. The CS decomposition of $W$ will now take the form in (\ref{eqn:CSD1}), with the matrix blocks $U$ and $X$ of size $3^{n-1} \times 3^{n-1}$ and blocks $V$ and $Y$ of size $3^{n-1}\cdot2 \times 3^{n-1}\cdot2 $. Repeating the partitioning process for the blocks $V$ and $Y$ with $m=3^{n-1}\cdot 2$ and $r=3^{n-1}$, and decomposing them with CSD followed by some matrix factoring will give rise to a decomposition of $W$ involving unitary blocks each of size $3^{n-1}$ as follows.

\begin{figure}\centerline{
 \Qcircuit @C.30em @R=.3em {
& &\qw &\qw  &\qw &\uctrld{1} &\qw &\qw &\qw &\qw & & & & &\cong & & & & & &\qw &\qw &\qw &\qw &\gate{+2} &\uctrld{2} &\gate{+2} &\uctrld{2} &\gate{+2} &\uctrld{2} &\qw &\qw \\
& & & &\gate{U_{0}} &\multigate{2}{M} & & & & & & & & & & & & \\
/ &\qw &\qw &\qw\qwx &\gate{U_{1}} &\ghost{M} &\qw &\qw &\qw &\qw & & & & & & & & & & &/ &\qw &\qw &\qw &\qw &\gate{U_{0}} &\qw &\gate{U_{1}} &\qw &\gate{U_{2}} &\qw &\qw\\
 & & &\qwx &\gate{U_{2}} &\ghost{M} \\}}\caption{\scriptsize{3-valued Quantum Multiplexer $M$ controlling the lower $(n-1)$ qutrits via the top qutrit. The slash symbol (/) represents $(n-1)$ qutrits on the second wire. The gates labeled +2 are shift gates, increasing the value of the qutrit by 2 mod 3, and the control $\diamondsuit$ turns on for input $\ket{2}$. Depending on the value of the top qutrit, one of $U_{t}$ is applied to the lower qutrits for $t\in\left\{0,1,2\right\}$.}}\label{fig:3-valued QMUX}
\end{figure}
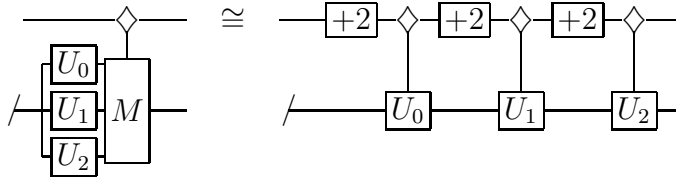

\begin{equation}\label{eqn:Ternary Decomposed CSD}
W=
ABC
\left(\begin{array}{ccc}
  C & -S & 0 \\ S & C & 0 \\  0 & 0 & I
\end{array}\right)
DEF
\end{equation}
with
\begin{equation}\label{eqn:ABC}
A =
\left(\begin{array}{ccc}
  X_{1} & 0 & 0 \\ 0 & X_{2} & 0 \\  0 & 0 & X_{3}
\end{array}\right),\hspace{0.25in}
B= \left(\begin{array}{ccc}
  I & 0 & 0 \\ 0 & C_{1} & -S_{1} \\  0 & S_{1} & C_{1}
\end{array}\right), \hspace{0.25in}
C= \left(\begin{array}{ccc}
  I & 0 & 0 \\ 0 & Z_{1} & 0 \\  0 & 0 & Z_{2}
\end{array}\right)
\end{equation}
\begin{equation}\label{eqn:DEF}
D= \left(\begin{array}{ccc}
Y_{1} & 0 & 0 \\ 0 & Y_{2} & 0 \\ 0 & 0 & Y_{3}
\end{array}\right), \hspace{0.25in}
E= \left(\begin{array}{ccc}
   I & 0 & 0 \\ 0 & C_{2} & -S_{2} \\  0 & S_{2} & C_{2}
\end{array}\right), \hspace{0.25in}
F= \left(\begin{array}{ccc}
 I & 0 & 0 \\ 0 & W_{1} & 0 \\  0 & 0 & W_{2}
\end{array}\right)
\end{equation}

We realize each block diagonal matrix in (\ref{eqn:ABC}) and (\ref{eqn:DEF}) as a 3-valued quantum multiplexer acting on $n$ qutrits of which the highest order qutrit is designated as the control qutrit. Depending on which of the values $\ket{0}$, $\ket{1}$, or $\ket{2}$ the control qutrit carries, the gate then performs either the top left block, the middle block, or the bottom right block respectively on the remaining $n-1$ qutrits. Figure~\ref{fig:3-valued QMUX} gives the layout for a $n$ qutrit quantum multiplexer realized in terms of \emph{Muthukrishnan-Stroud} (MS) gates. The MS gate is a $d$-valued generalization of the controlled-not (CNOT) gate from 2-valued quantum logic, and allows for control of one qudit by the other via the highest value of a $d$-valued quantum system, which in the 3-valued case is 2~\cite{MuthuStroud:04}.

\begin{figure}\centerline{
 \Qcircuit @C=1em @R=.8em {
& &\multigate{1}{R_{x}} &\qw & &\cong & & &\gate{R_{x}^{\theta_{0}}} &\gate{R_{x}^{\theta_{1}}} &\gate{R_{x}^{\theta_{2}}} &\qw &\dots & &\gate{R_{x}^{\theta_{(3^{n-1})-1}}} &\qw \\
& /&\ghost{R_{x}} &\qw & & & & &\ctrlo{-1} &\ctrlo{-1} &\ctrlo{-1} &\qw &\dots & &\uctrld{-1} &\qw  \\
& & & & & & & &\ctrlo{-1}  &\ctrlo{-1} &\ctrlo{-1} &\qw &\dots & &\uctrld{-1} &\qw \\
& & & & & & & &\vdots &\vdots &\vdots & &\vdots & &\vdots\\
& & & & & & & &\ctrlo{1} &\ctrlo{1} &\ctrlo{1} &\qw &\dots & &\uctrld{1} &\qw\\
& & & & & & & &\ctrlo{-1} &\ctrl{-1} &\uctrld{-1} &\qw &\dots & &\uctrld{-1} &\qw\\}}\caption{\scriptsize{A uniformly $(n-1)$-controlled $R_{x}$ rotation. The lower $(n-1)$ qutrits are the control qutrits represented on the left hand side  by the symbol / on the second wire. The controls $\circ$, $\bullet$, and $\diamond$ turn on for inputs $\ket{0}$, $\ket{1}$, and $\ket{2}$ respectively. It requires $3^{n-1}$ one qutrit controlled gates to implement a uniformly $(n-1)$-controlled $R_{x}$ or $R_{z}$ rotation. }}\label{fig:(n-1)-unifo control Rx}
\end{figure}
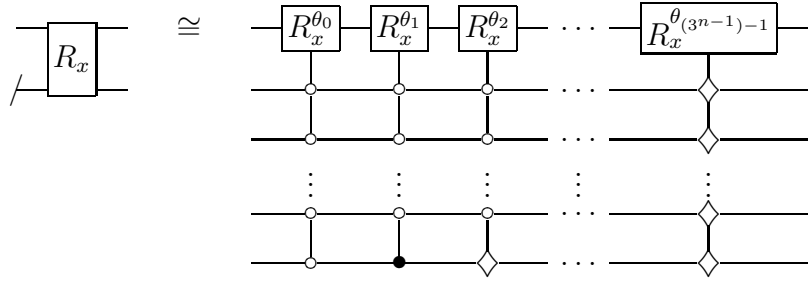
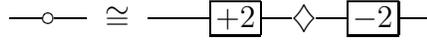
\begin{figure}\centerline{
 \Qcircuit @C1em @R=.8em {
&\controloo &\qw &\cong & &\qw &\gate{+2} &\uniformcontroldd &\gate{-2} &\qw  }}
\caption{\scriptsize{A control by the value 0 (mod 3) realized in terms of control by the highest value 2 (mod 3).}}\label{fig:0-control mod 3}
\end{figure}
\hfill
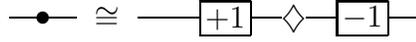
\begin{figure}\centerline{
 \Qcircuit @C1em @R=.8em {
&\controll &\qw &\cong & &\qw &\gate{+1} &\uniformcontroldd &\gate{-1} &\qw  }}
\caption{\scriptsize{A control by the value 1 (mod 3) realized in terms of control by the highest value 2 (mod 3).}}\label{fig:1-control mod 3}
\end{figure}

The cosine-sine matrices are realized as the uniformly $(n-1)$-controlled $R_{x}$ and $R_{z}$ rotations in $\textbf{R}^{3}$. Similar to the 2-valued case, each $R_{x}$ and $R_{z}$ rotation is composed of a sequence of $(n-1)$-fold controlled gates $R^{\theta_{i}}_{x}$ or $R^{\phi_{i}}_{z}$, where 
\begin{equation}\label{eqn:R_{x}}
R^{\theta_{i}}_{x} = \left(\begin{array}{ccc}
 1 & 0 & 0 \\ 0 & \cos\theta_{i} & -\sin\theta_{i}  \\ 0 &  \sin\theta_{i}&  \cos\theta_{i} \\
\end{array}\right),\hspace{0.25in}
R^{\phi_{i}}_{z} = \left(\begin{array}{ccc}
\cos\phi_{i} & -\sin\phi_{i}  & 0 \\  \sin\phi_{i} &  \cos\phi_{i} & 0 \\ 0 & 0 & 1
\end{array}\right).
\end{equation}
Each $R^{\theta_{i}}_{x}$ or $R^{\phi_{i}}_{z}$ operator is applied to the top most qutrit, with the value of the angles $\theta_{i}$ and $\phi_{i}$ determined by the $(n-1)$ basis state configurations of the control qutrits. A uniformly controlled $R_{x}$ gate is shown in figure~\ref{fig:(n-1)-unifo control Rx}. Figures~\ref{fig:0-control mod 3} and~\ref{fig:1-control mod 3} explain the method to create controls of maximum value. The value of the control qubit is always restored in figures~\ref{fig:0-control mod 3} and~\ref{fig:1-control mod 3}.

\section{Synthesis of Hybrid and $d$-valued Quantum Logic Circuits}\label{sect:hybrid}
It is evident from the 2 and 3-valued cases above that the CSD method of synthesis is of a general nature and can be extended to synthesis of $d$-valued gates acting on $n$ qudits. In fact, it can be generalized for synthesis of hybrid $n$ qudit gates. We propose that a $(d_{1}d_{2}\dots d_{n}) \times (d_{1}d_{2}\dots d_{n})$ block diagonal unitary matrix be regarded as a quantum multiplexer for an $n$ qudit hybrid quantum state space $\mathcal{H}=\mathcal{H}_{d_{1}} \otimes \mathcal{H}_{d_{2}} \otimes \dots \otimes \mathcal{H}_{d_{n}}$, where $\mathcal{H}_{d_{i}}$ is the state space of the $i$ qudit. 

\begin{figure}\centerline{
 \Qcircuit @C=.30em @R=.5em {
& \qw &\qw &\uctrl{1} &\qw & & & &\cong & & & & &\qw &\gate{+(d_{i}-1)} &\uctrl{2} &\gate{+(d_{i}-1)} &\uctrl{2} &\gate{+(d_{i}-1)} &\qw \qw &\qw & & & &\dots & & & &\gate{+(d_{i}-1)} &\uctrl{2} &\qw \\ 
& &\gate{X_{0}} &\multigate{4}{M} & &  & & & & & & & & & & & & & & & & & & &\vdots  \\
&\qwx &\gate{X_{1}} &\ghost{M} & & & & & & & & & / &\qw &\qw &\gate{X_{0}} &\qw &\gate{X_{1}} &\qw &\qw & & & & &\dots & & & &\qw &\gate{X_{d_{i}-1}} &\qw \\
/ &\qw\qwx &\vdots  &\gost{M} &\qw &\qw & & & & & & & & & & &  & & & & & & & & \\
&\qwx & &  & & & & & & & & & & & & & & & & &   \\
&\qwx &\gate{X_{d_{i}-1}} &\ghost{M} & & & & & & &  \\}}
\caption{\scriptsize{An $n$ qudit hybrid quantum multiplexer, here realized in terms of Muthukrishnan-Stroud ($d$-valued controlled) gates. The top qudit has dimension $d_{i}$ and controls the remaining $(n-1)$ qudits of possibly distinct dimensions which are represented here by the symbol (/). The control $\oslash$ turns on for input value $\ket{d_{i}-1}$ mod $d_{i}$ of the controlling signal coming from the top qudit.The gates $+(d_{i}-1)$ shift the values of control qudit by $(d_{i}-1)$ mod $d_{i}$.}}\label{fig:hybrid QMUX}
\end{figure}
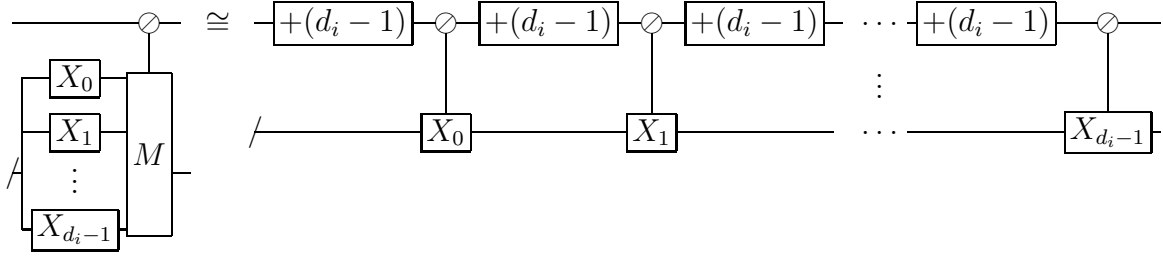

Moreover, consider a cosine-sine matrix of size $(d_{1}d_{2}\dots d_{n}) \times (d_{1}d_{2}\dots d_{n})$ of the form
\begin{equation}
\left(\begin{array}{cccc} I_{p} & 0 & 0 & 0\\ 0 & C & -S
& 0
\\ 0 & S & C  & 0 \\ 0 & 0 & 0 & I_{q}
\end{array}\right)
\end{equation}
with $I_{p}$ and $I_{q}$ both some appropriate sized identity matrices, $C$ = diag$(\cos \theta_{1}, \\ \cos\theta_{2}, \dots, \cos\theta_{t})$ and $S$ = diag$(\sin \theta_{1}$, $\sin \theta_{2},\dots,\sin \theta_{t})$ such that $\sin^{2}\theta_{i}+\cos^{2}\theta_{i}=1$ for some $\theta_{i}$ with $1 \leq i \leq t$, and $p+q+2t=(d_{1}d_{2}\dots d_{n})$. We regard this matrix as a \emph{uniformly controlled Givens rotation} matrix, a generalization of the $R_{y}$, $R_{x}$, and $R_{z}$ rotations of the 2 and 3-valued cases. A Givens rotation matrix has the general form
\begin{equation}\label{eqn:givens matrix}
G_{(i,j)}^{\theta} = \left(\begin{array}{ccccccc}
1 & \dots & 0 & \dots & 0 & \dots & 0 \\
\vdots & \ddots & \vdots &  & \vdots & & \vdots \\
0 & \dots & \cos\theta & \dots & -\sin\theta & \dots & 0 \\
\vdots &  & \vdots  & \ddots &  \vdots &  & \vdots \\
0 & \dots & \sin\theta & \dots & \cos\theta & \dots & 0 \\
\vdots & & \vdots & & \vdots & \ddots & \vdots \\
0 & \dots & 0 & \dots & 0 & \dots & 1\\
\end{array}\right)
\end{equation}
where the cosine and sine values reside in the intersection of the $i$-th and $j$-th rows and columns, and all other diagonal entries are 1~\cite{Golub:89}. Hence, a Givens rotation matrix corresponds to a rotation by some angle $\theta$ in the $ij$-th hyperplane.

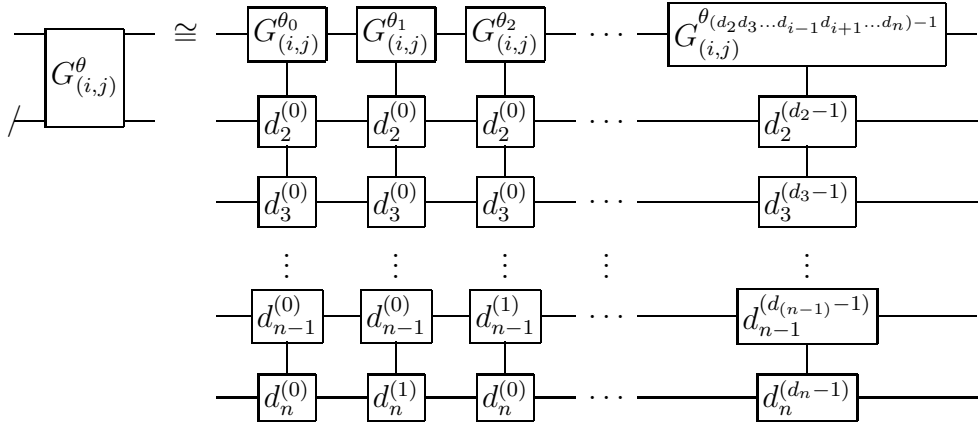
\begin{figure}\centerline{
 \Qcircuit @C=1em @R=1em {
& & &\multigate{1}{G_{(i,j)}^{\theta}} &\qw &\cong & &\gate{G_{(i,j)}^{\theta_{0}}} &\gate{G_{(i,j)}^{\theta_{1}}} &\gate{G_{(i,j)}^{\theta_{2}}} &\qw &\dots & &\gate{G_{(i,j)}^{\theta_{(d_{2}d_{3}\dots d_{i-1}d_{i+1}\dots d_{n})-1}}} &\qw \\
& &/ &\ghost{G_{(i,j)}} &\qw  & & &\gate{d_{2}^{(0)}}\qwx &\gate{d_{2}^{(0)}}\qwx &\gate{d_{2}^{(0)}}\qwx &\qw &\dots & &\gate{d_{2}^{(d_{2}-1)}}\qwx &\qw  \\
& & & &  &  & &\gate{d_{3}^{(0)}}\qwx &\gate{d_{3}^{(0)}}\qwx &\gate{d_{3}^{(0)}}\qwx &\qw &\dots & &\gate{d_{3}^{(d_{3}-1)}}\qwx &\qw \\
& & & & & & &\vdots &\vdots &\vdots & &\vdots & &\vdots\\
& & & &  & & &\gate{d_{n-1}^{(0)}} &\gate{d_{n-1}^{(0)}} &\gate{d_{n-1}^{(1)}} &\qw &\dots & &\gate{d_{n-1}^{(d_{(n-1)}-1)}} &\qw\\
& & & & & & &\gate{d_{n}^{(0)}}\qwx &\gate{d_{n}^{(1)}}\qwx &\gate{d_{n}^{(0)}}\qwx &\qw &\dots & &\gate{d_{n}^{(d_{n}-1)}}\qwx &\qw\\
}}
\caption{\scriptsize{A hybrid uniformly $(n-1)$-controlled Givens rotation. The lower $(n-1)$ qudits of dimensions $d_{2},d_{3}, \dots , d_{i-1},d_{i+1},\dots , d_{n}$, respectively, are the control qudits, and the top is the target qudit of dimension $d_{i}$. The control gate $d_{l}^{(k)}$ turns on whenever the control qudit of dimension $d_{l}$ takes on the value $k$ (mod) $d_{l}$.}}\label{fig:hybrid (n-1)-unifo control Givens}
\end{figure}
\hfill
\begin{figure}\centerline{
 \Qcircuit @C1em @R=.8em {
&\gate{d_{l}^{(k)}} &\qw &\cong & &\qw &\gate{+(d_{l}-1)} &\gate{d_{l}} &\gate{-(d_{l}-1)} &\qw  }}
\caption{\scriptsize{A control by the value $k$ (mod $d_{l})$ realized via control by the highest value $(d_{l}-1)$ (mod $d_{l}$).}}\label{fig:1-control mod dl}
\end{figure}

Based on the preceding discussion, we give in theorem 5.1.1 below an iterative CSD method for synthesizing a $n$ qudit hybrid quantum circuit by decomposing the corresponding unitary matrix of size $(d_{1}d_{2}\dots d_{n}) \times (d_{1}d_{2}\dots d_{n})$ in terms of quantum multiplexers and uniformly controlled Givens rotations. As a consequence of this theorem, we give in lemma 5.1.1 a CSD synthesis of a quantum quantum logic circuit with corresponding unitary matrix of size $d^{n} \times d^{n}$. The synthesis methods given above for 2-valued and 3-valued circuits may then be treated as special cases of the former.

\subsection{Hybrid Quantum Logic Circuits}
Consider a hybrid quantum state space of a $n$ qudits, $\mathcal{H}=\mathcal{H}_{d_{1}} \otimes \mathcal{H}_{d_{2}} \otimes \dots
\otimes \mathcal{H}_{d_{n}}$, where each qudit may be of distinct $d$-valued dimension $d_{i}$, $0 \leq i \leq n$. Since a qudit in $\mathcal{H}$ is a column vector of length $d_{1}d_{2} \ldots d_{n}$, a quantum logic gate acting on such a vector is a $(d_{1}d_{2} \ldots d_{N}) \times (d_{1}d_{2} \ldots d_{n})$ unitary matrix $W$. We will decompose $W$, using CSD iteratively, from the level of $n$ qudits to $(n-1)$ qudits in terms of quantum multiplexers and uniformly controlled Givens rotations. However, since the $d$-valued dimension may be different for each qudit, the block matrices resulting from the CS decomposition may not be of the form $d^{n-1} \times d^{n-1}$ for some $d$. Therefore, we proceed by choosing one of the qudits, $c_{d_{i}}$ of dimension $d_{i}$, to be the control qudit and order of the basis of $\mathcal{H}$ in such a way that $c_{d_{i}}$ is the highest order qudit. We will decompose $W$ with respect to $c_{d_{i}}$ so that the resulting quantum multiplexers are controlled by $c_{d_{i}}$ and the uniformly controlled Givens rotations control $c_{d_{i}}$ via the remaining $(n-1)$ qudits. We give the synthesis method in theorem 5.1.1.

\emph{\textbf{Theorem 5.1.1:} Let $W$ be an $M \times M$ unitary matrix, with $M=d_{1}d_{2}\ldots d_{n}$, acting as a quantum logic gate on a quantum hybrid state space $\mathcal{H}=\mathcal{H}_{d_{1}} \otimes \mathcal{H}_{d_{2}} \otimes \dots \otimes \mathcal{H}_{d_{n}}$ of $n$ qudits. Then $W$ can be synthesized with respect to a control qudit $c_{d_{i}}$ of dimension $d_{i}$, having the highest order in $\mathcal{H}$, iteratively from level $n$ to level $(n-1)$ in terms of quantum multiplexers and uniformly controlled Givens rotations}. 

\textbf{Proof}.
 
\textbf{Step 1.} At level $n$, identify a control qudit $c_{d_{i}}$ of dimension $d_{i}$. Reorder the basis of $\mathcal{H}$ so that $c_{d_{i}}$ is the highest order qudit and the new state space isomorphic to $\mathcal{H}$ is $\mathcal{\bar{H}}=\mathcal{H}_{d_{i}} \otimes \mathcal{H}_{d_{2}} \otimes \dots
\otimes \mathcal{H}_{d_{1}} \otimes \dots \otimes \mathcal{H}_{d_{n}}$.

If we choose values for the CSD parameters $m$ and $r$ as $m=\left(d_{1}d_{2}\dots d_{n}\right)$ and $r=\left(d_{1}d_{2}\ldots d_{i-1}d_{i+1}\dots d_{n}\right)$, then $m-r=d_{1}\ldots d_{i-1}d_{i+1}\dots d_{n}(d_{i}-1)$. Decomposing $W$ by CSD, we get the form in (\ref{eqn:CSD1}) with the matrix blocks  $U$ and $X$ of size $r \times r$ and blocks $V$ and $Y$ of size $(m-r) \times (m-r)$. Should $m-r$ not have the factor $(d_i-1)$, we would achieve the desired decomposition of $W$ from level of $n$ qudits to the level of $(n-1)$ qudits in terms of block matrices of size $r \times r$. The task therefore is to divide out the factor $(d_{i}-1)$ from $m-r$ by an iterative \emph{lateral decomposition} described below, that uses the CSD to cancel $(d_{i}-1)$ from $m-r$ at each iteration level leaving only blocks of size $r \times r$. 

For step 2 of the proof below, we will say that a matrix with $k$ rows and $k$ columns has size $k$ instead of $k \times k$. 

\textbf{Step 2.} \emph{Iterative Lateral Decomposition}: For the unitary matrix $W$ of size $M$, we define the $j$-th lateral decomposition of $W$ as the CS decomposition of all block matrices of size other than $r$ that result from the $(j-1)$-st lateral decomposition of $W$:

\emph{
For $0\leq j \leq (d_{i}-2)$, set \\
$m_{0}=\left(d_{1}d_{2} \ldots d_{n}\right)$ \\
$r_{0}=\left(d_{1}d_{2}\ldots d_{i-1}d_{i+1}\dots d_{n}\right)$ \\
If $j=0$ \\
Apply CSD to W\\
Else set \\
$m_{j}=m_{0}-j\cdot r_{0}$ \\
$r_{j}=r_{0}$ \\
$m_{j}-r_{j}=m_{0}-(j+1)r_{0}$\\
=$\left(d_{1}d_{2}\ldots d_{i-1}d_{i+1}\dots d_{n}\right)[d_{i}-(j+1)]$\\
$m_{j}-2r_{j}=m_{0}-(j+2)r_{0}$\\
=$\left(d_{1}d_{2}\ldots d_{i-1}d_{i+1}\dots d_{n}\right)[d_{i}-(j+2)]$\\
Apply CSD to matrix blocks of size other than $r_{0}$ from step $j-1$\\
End If\\
End For}.

When $j=0$, we call the resulting 0-th lateral decomposition the \emph{global decomposition}. Note that if $d_{i}=2$, then the algorithm for the lateral decomposition stops after the global decomposition. This suggests that whenever feasible, the control system in the quantum circuit should be 2-valued so as to reduce the number of iterations . Below we give a matrix description of the algorithm.

For $j=0$, the $0$-th lateral decomposition of $W$ will just be the CS decomposition of $W$.
\begin{equation}\label{eqn:lateral1}
 W=A_{0}^{(0)}B_{0}^{(0)}C_{0}^{(0)}
\end{equation}
where
$$
A_{0}^{(0)}=\left(\begin{array}{cc}
 U^{(0)}_{0} & 0 \\ 0 & V^{(0)}_{0}
\end{array}\right),
B_{0}^{(0)}=\left(\begin{array}{ccc}
  C^{(0)}_{0} & -S^{(0)}_{0} & 0 \\ S^{(0)}_{0} & C^{(0)}_{0} & 0 \\ 0 & 0 & I_{m_{0}-2r_{0}}
\end{array}\right)
C_{0}^{(0)}=\left(\begin{array}{cc}
  X^{(0)}_{0} & 0 \\  0 & Y^{(0)}_{0}
\end{array}\right)
$$
with $U^{(0)}_{0}$, $X^{(0)}_{0}$, $C^{(0)}_{0}$, and $S^{(0)}_{0}$ all of the
desired size $r_{0}$, while $V^{(0)}_{0}$ and $Y^{(0)}_{0}$ are of size
$m_{0}-r_{0}$. The superscripts label the iteration step, in this case $j=0$. The
subscript is used to distinguish between the various matrix blocks $U, V, X, Y, C,
S$, that occur at the various levels of iteration. The 0-th lateral decomposition in the form from equation
(\ref{eqn:lateral1}) is called the \emph{global decomposition} of $W$. 

For $j=1$, we perform lateral decomposition on the blocks $V^{(0)}_{0}$ and $Y^{(0)}_{0}$ of the block matrices $A_{0}^{(0)}$ and $C_{0}^{(0)}$ respectively, the only blocks of size other than $r_{0}$ resulting from the $0$-th lateral decomposition given in (\ref{eqn:lateral1}). In both cases, set $ m_{1}=m_{0}-r_{0}$ and $r_{1}=r_{0}$ so that $m_{1}-r_{1}=m_{0}-2r_{0}$. For $V^{(0)}_{0}$ this gives the decomposition
\begin{equation}\label{eqn:lateral2}
A_{0}^{(0)}=\left(\begin{array}{cc} U^{(0)}_{0} & 0 \\ 0 &
\left(\begin{array}{cc} U^{(1)}_{0} & 0 \\
0 & V^{(1)}_{0} \end{array}\right)
\left(\begin{array}{ccc} C_{0}^{(1)} & -S_{0}^{(1)} & 0 \\ S_{0}^{(1)} & C_{0}^{(1)}  & 0 \\
0 & 0 & I_{m_{0}-3r_{0}} \end{array}\right) \left(\begin{array}{cc} X^{(1)}_{0} &
0 \\ 0 & Y^{(1)}_{0}
\end{array}\right)
\end{array}\right)
\end{equation}
with $U^{(1)}_{0}, X^{(1)}_{0}, C_{0}^{(1)}$ and $S_{0}^{(1)}$ all of size
$r_{0}$, and $V^{(1)}_{0}$ and $Y^{(1)}_{0}$ of size $m_{1}-r_{1}$. All three matrices residing in the lower block diagonal of the matrix (\ref{eqn:lateral2}) are the same size. Therefore, by introducing identity matrices of size $r_{0}$ and factoring out at the matrix block level, $A_{0}^{(0)}$ will be updated to
\begin{equation}
A_{0}^{(0)}=A_{0}^{(1)}B_{0}^{(1)}C_{0}^{(1)}
\end{equation}
where
$$
A_{0}^{(1)}=\left(\begin{array}{ccc} U^{(0)}_{0} & 0 & 0 \\ 0 & U^{(1)}_{0} &  0 \\ 0 & 0 &
V^{(1)}_{0}
\end{array}\right),
B_{0}^{(1)}=\left(\begin{array}{cccc} I_{r_{0}} & 0 & 0 & 0\\ 0 & C_{0}^{(1)} & -S_{0}^{(1)}
& 0
\\ 0 & S_{0}^{(1)} & C_{0}^{(1)}  & 0 \\ 0 & 0 & 0 & I_{m_{0}-3r_{0}}
\end{array}\right),
C_{0}^{(1)}=\left(\begin{array}{ccc} I_{r_{0}} & 0 & 0 \\ 0 & X^{(1)}_{0}& 0 \\  0 & 0 &
Y^{(1)}_{0}
\end{array}\right)
$$
A similar lateral decomposition of the block $Y^{(0)}_{0}$ will update $C_{0}^{(0)}$ in (\ref{eqn:lateral1}) to
\begin{equation}
C_{0}^{(0)}=A_{1}^{(1)}B_{1}^{(1)}C_{1}^{(1)}
\end{equation}
where
$$
A_{1}^{(1)}=\left(\begin{array}{ccc} X^{(0)}_{0} & 0 & 0 \\ 0 & U^{(1)}_{1} &  0 \\ 0 & 0 &
V^{(1)}_{1} \end{array}\right),
B_{1}^{(1)}=\left(\begin{array}{cccc} I_{r_{0}} & 0 & 0 & 0\\ 0 & C_{1}^{(1)} & -S_{1}^{(1)}
& 0
\\ 0 & S_{1}^{(1)} & C_{1}^{(1)}  & 0 \\ 0 & 0 & 0 & I_{m_{0}-3r_{0}}
\end{array}\right),
C_{1}^{(1)}=\left(\begin{array}{ccc} I_{r_{0}} & 0 & 0 \\ 0 & X^{(1)}_{1} & 0 \\  0 & 0 &
Y^{(1)}_{1}
\end{array}\right)
$$

For iteration $j \neq 0$, perform lateral decomposition on the total
$2^{j}$ blocks $V^{(j-1)}_{k}$, $Y^{(j-1)}_{k}$, where $0 \leq k \leq (j-1)$,
that occur in the global decomposition at the end of iteration $(j-1)$. For each
$V^{(j-1)}_{k}$, $Y^{(j-1)}_{k}$, set $r_{j}=r_{0}$, $m_{j}=m_{j-1}-r_{j-1}=m_{0}-(j+1)r_{0}$.
For each $V^{(j-1)}_{k}$, the lateral decomposition at level $j$ will give the following
\begin{equation}\label{eqn:lateral3}
A_{k^{\prime}}^{(j)}=\left(\begin{array}{cc} \Delta^{(j-1)} & 0 \\ 0 &
\left(\begin{array}{cc} U^{(j)}_{k^{\prime}} & 0 \\
0 & V^{(j)}_{k^{\prime}} \end{array}\right)
\left(\begin{array}{ccc} C^{(j)}_{k^{\prime}} & -S^{(j)}_{k^{\prime}} & 0 \\ S^{(j)}_{k^{\prime}}& C^{(j)}_{k^{\prime}}
& 0 \\
0 & 0 & I_{m_{0}-(j+2)r_{0}} \end{array}\right) \left(\begin{array}{cc}
X^{(j)}_{k^{\prime}}& 0 \\ 0 & Y^{(j)}_{k^{\prime}}
\end{array}\right)
\end{array}\right)
\end{equation}
where the $\Delta^{(j-1)}$ is the block diagonal matrix of size of $j\cdot
r_{0}$ arising from the lateral decomposition in the previous $(j-1)$ steps. The
blocks $U^{(j)}_{k^{\prime}}$, $X^{(j)}_{k^{\prime}}$, $C^{(j)}_{k^{\prime}}$ and
$S^{(j)}_{k^{\prime}}$ are all of size $r_{0}$, for $0\leq k^{\prime} \leq j$.
The blocks $V^{(j)}_{k^{\prime}}$ and $Y^{(j)}_{k^{\prime}}$ are of size
$m_{j}-r_{j}$. The three matrices residing in the lower block diagonal of the matrix (\ref{eqn:lateral3}) are all of same size. Therefore, by introducing identity matrices of size $j\cdot r_{0}$ and factoring out at the block level, $A_{k^{\prime}}^{(j)}$ will be updated to
$$
A_{k^{\prime}}^{(j)}=A_{k^{\prime}}^{(j+1)}B_{k^{\prime}}^{(j+1)}C_{k^{\prime}}^{(j+1)}
$$
where
$$
A_{k^{\prime}}^{(j)}=\left(\begin{array}{ccc} \Delta^{(j-1)} & 0 & 0 \\ 0 & U^{(j)}_{k^{\prime}} &  0 \\
0 & 0 & V^{(j)}_{k^{\prime}}
\end{array}\right),
B_{k^{\prime}}^{(j)}=\left(\begin{array}{cccc} I_{j\cdot r_{0}} & 0 & 0 & 0\\ 0 &
C^{(j)}_{k^{\prime}} &
 -S^{(j)}_{k^{\prime}} & 0
\\ 0 &  S^{(j)}_{k^{\prime}} & C^{(j)}_{k^{\prime}}  & 0 \\ 0 & 0 & 0 & I_{m_{0}-(j+2)r_{0}}
\end{array}\right)
C_{k^{\prime}}^{(j)}=\left(\begin{array}{ccc} I_{j\cdot r_{0}} & 0 & 0 \\ 0 & X^{(j)}_{k^{\prime}} & 0 \\
0 & 0 & Y^{(j)}_{k^{\prime}}
\end{array}\right)
$$
For the next iteration, set $k=k^{\prime}$ and iterate. Upon completion of the lateral decomposition, repeat steps 1 and 2 for the synthesis of the circuit for the remaining $(n-1)$ qudits, with the restriction that each gate in the remaining circuit be decomposed with respect to the same control qudit identified in step 1. 

Since the basis for $\mathcal{H}$ was reordered in the beginning so that the control qudit was of the highest order, the block diagonal matrices with all blocks of size $r_{0} \times r_{0}$ are interpreted as quantum multiplexers and the cosine-sine matrices are interpreted as uniformly controlled Givens rotations. In figures~\ref{fig:hybrid QMUX} and \ref{fig:hybrid (n-1)-unifo control Givens}, we present the circuit diagrams of a hybrid quantum multiplexer and a uniformly controlled Givens rotation, respectively. A uniformly controlled Givens rotation matrix on $n$ qudits can be realized as the composition of various $(n-1)$-fold controlled Givens rotation matrices, $G_{(i,j)}^{\theta_{k}}$, acting on the top most qudit of the circuit with the angle of rotation depending on the basis state configuration, in their respective dimensions, of the lower $(n-1)$ qudits.

\subsection{d-valued Quantum Logic Circuits}\label{sect:d-valued}
Given the hybrid $n$ qudit synthesis, the case of $d$-valued synthesis becomes a special case of the former since by setting all $d_{i}=d$, the state space $\mathcal{H}=\mathcal{H}_{d_{1}} \otimes \mathcal{H}_{d_{2}} \otimes \dots \otimes \mathcal{H}_{d_{n}}$ reduces to the state space $\mathcal{H}^{\otimes n}$. Unitary operators acting on the states in $\mathcal{H}^{\otimes n}$ are unitary matrices of size $d^{n} \times d^{n}$. We give the following result for $d$-valued synthesis.

\emph{\textbf{Lemma 5.1.1:} A $d$-valued $n$ qudit quantum logic gate can be synthesized in terms of quantum multiplexers and uniformly controlled Givens rotations}.

\textbf{Proof}: Since all the qudits are of the same dimension, there is no need to choose a control qudit. In the proof of theorem 5.1.1, set $d_{i}=d$ for all $i$. Then $M=d_{1}d_{2}\dots d_{n}=d^{n}$. For iteration $j=0$ of the lateral decomposition, set $m_{0}=d^{n}$, $r_{0}=d^{n-1}$, so that $m_{0}-r_{0}=d^{n-1}(d-1)$. For $0 \leq j \leq (d-2)$, set $r_{j}=r_{0}=d^{n-1}$, and $m_{j}=m_{j-1}-r_{j-1}=d^{n-1}(d-(j+1))$.

For the $d$-valued case, we note that there are a total of $d^{n-1}(2^{d-1}-1)$ one qudit Givens rotations in the circuit at the $(n-1)$ level, each arising from the $\sum_{i=0}^{(d-2)}2^{i}=2^{d-1}-1$ uniformly controlled Givens rotations in the CS decomposition of an $n$ qudit gate. Moreover, in each uniformly controlled Givens rotation, there are $(n-1)d^{n-1}$ control symbols of which $(n-1)d^{n-2}$ correspond to control by the highest value of $d-1$. The latter controls do not require shift gates around them to increase the value of the signal qudit to $d-1$. Hence, there are $(n-1)d^{n-1}-(n-1)d^{n-2}=(n-1)(d^{n-1}-d^{n-2})$ control symbols that correspond to control by values other than $d-1$ and therefore need two shift gates (fig. 11) around them. This gives the total number of one qudit shift gates in each uniformly controlled rotation to be $2(n-1)(d^{n-1}-d^{n-2})$, whereby the total number of one qudit shifts and Givens rotations in the circuit at the $(n-1)$ level is $2(n-1)(d^{n-1}-d^{n-2})(2^{d-1}-1)+d^{n-1}(2^{d-1}-1)=(2^{d-1}-1)\left[2(n-1)(d^{n-1}-d^{n-2})+d^{n-1}\right]$.

There are $2^{d-1}$ quantum multiplexers in the decomposition, each consisting of a total of $2d$ shift and controlled gates. Hence, there are a total of  $d\cdot 2^d$ one qudit and controlled gates in the $(n-1)$ level circuit. This gives a total, worst case, one qudit and controlled gate count in the circuit at level $(n-1)$ to be $(2^{d-1}-1)\left[2(n-1)(d^{n-1}-d^{n-2})+d^{n-1}\right]+ d\cdot 2^d$.

\section{Conclusion}
We have shown that the method of CS decomposition of unitary matrices used in 2-valued and 3-valued quantum logic synthesis is a special case of a general synthesis method based on the CSD. We give an algorithm for this general method that allows us to synthesize $n$ qudit hybrid and $d$-valued quantum logic circuits in terms of quantum multiplexers and uniformly controlled Givens rotations.

\section{Acknowledgments}
F.S. Khan is grateful to Steve Bleiler for discussions and suggestions. The Quantum Circuit diagrams were all drawn in \LaTeX \ using Q-circuit available at {http://info.phys.unm.edu/Qcircuit/}.

\end{document}